\documentclass[a4paper, 11pt]{amsart}
\usepackage{amsmath, amssymb, amsthm}	
\usepackage[foot]{amsaddr}				
\usepackage{braket, stmaryrd, MnSymbol,ascmac,mathrsfs}
\usepackage{cite}

\calclayout

\newcommand{\pr}[1]{#1^{\prime}}

\newcommand{\dg}[1]{#1^{\dag}}
\newcommand{\del}{\partial}

\newcommand{\mfrak}[1]{\mathfrak{#1}}
\newcommand{\mcal}[1]{\mathcal{#1}}
\newcommand{\mbb}[1]{\mathbb{#1}}
\newcommand{\mrm}[1]{\mathrm{#1}}

\newcommand{\scr}[1]{\mathscr{#1}}
\newcommand{\what}[1]{\widehat{#1}}
\newcommand{\no}[1]{:\hspace{-3pt} #1\hspace{-3pt}:\hspace{3pt}}

\makeatletter
    
    \@addtoreset{equation}{section}
\makeatother

\title[SLE-type growth processes corresponding to WZW theories]{Schramm--Loewner-evolution-type growth processes corresponding to Wess--Zumino--Witten theories}
\author{Shinji Koshida}
\address{Department of Basic Science, The University of Tokyo}
\email{koshida@vortex.c.u-tokyo.ac.jp}

\begin{document}

\begin{abstract}
A group theoretical formulation of Schramm--Loewner-evolution-type growth processes corresponding to
Wess--Zumino--Witten theories is developed that makes it possible to construct stochastic differential equations
associated with more general null vectors than the ones considered in the most fundamental example in [Alekseev {\it et al.}, Lett. Math. Phys. {\bf 97}, 243-261 (2011)].
Also given are examples of Schramm--Loewner-evolution-type growth processes associated with
null vectors of conformal weight $4$ in the basic representations of $\what{\mfrak{sl}}_{2}$ and $\what{\mfrak{sl}}_{3}$.
\end{abstract}

\maketitle

\section{Introduction}
Two-dimensional conformal field theories (CFTs) \cite{BelavinPolyakovZamolodchikov1984} have been proven to be 
powerful tools for analyzing a vast variety of models in statistical physics at criticality \cite{DiFrancescoMathieuSenechal1997}.
CFTs are distinguished from other quantum field theories by the property that several correlation functions in a CFT can be computed exactly
owing to its infinite-dimensional symmetry.
Not only can a bulk system without boundaries be probed but so can one under certain boundary conditions
by means of the framework of boundary conformal field theories (BCFTs) \cite{Cardy1986,Cardy1989}.
In a BCFT, a partition function under given boundary conditions is defined by a correlation function of boundary operators,
each of which changes boundary conditions.
In this picture, a partition function of a BCFT gives a prediction for a cluster interface evolving from one boundary point to another.
One of the milestones is Cardy's formula \cite{Cardy1992}, later proved by Smirnov \cite{Smirnov2001}, 
which gives the crossing probability in two-dimensional critical percolation computed as partition functions of a BCFT.

Another framework describing a cluster interface is
the notion of Schramm--Loewner evolution (SLE), which was introduced in \cite{Schramm2000} as a subsequential scaling limit
of loop-erased random walks and uniform spanning trees.
Schramm defined in \cite{Schramm2000} two types of SLEs, radial and chordal,
but in this paper, we only consider chordal SLE and simply call it SLE.
SLE labeled by a positive real number $\kappa > 0$ is a collection of properly normalized
conformal maps $\{g_{t}(z)\}_{t\ge 0}$ satisfying the following ordinary differential equation:
\begin{equation}
	\label{eq:sle}
	\del_{t}g_{t}(z)=\frac{2}{g_{t}(z)-B_{t}},\ \ g_{0}(z)=z.
\end{equation}
Here $B_{t}$ is Brownian motion of variance $\kappa$ on the real axis.
Following its introduction, many properties of SLE have been unveiled mainly in a probability theoretical manner \cite{RohdeSchramm2005, Lawler2004}.
In particular, it has been found that SLE emerges in scaling limits of several two-dimensional lattice models including the spin Ising model \cite{ChelkakDuminil-CopinHonglerKemppainenSmirnov2014, Lawler2005, Smirnov2006}.

Because CFT and SLE are two different frameworks used to treat the same critical systems, they should be connected in some sense.
There are several approaches to relate these two notions under the name of SLE/CFT correspondence.
In works by Friedrich, Kalkkinen, and Kontsevich \cite{FriedrichKalkkinen2004,Friedrich2004,Kontsevich2003},
the SLE-type probability measure was constructed as a section of the determinant bundle over
the moduli space of Riemann surfaces.
Recently, Dub{\'e}dat \cite{Dubedat2015a,Dubedat2015b} also constructed SLE-type probability measures
by means of the localization technique and identified its partition function with a highest weight vector in a highest weight representation
of the Virasoro algebra. 

In this paper, we follow the {\it group theoretical} formulation of SLE/CFT correspondence originated by 
Bauer and Bernard \cite{BauerBernard2002,BauerBernard2003a,BauerBernard2003b}.
They established the relation between SLE and CFT via a random process on an infinite-dimensional Lie group.
Their strategy was to consider a random process $\scr{G}_{t}$ on the {\it lower Borel subgroup} of the Virasoro group,
of which the Lie algebra is the Virasoro algebra, under the initial condition that $\scr{G}_{0}$ is the unit element.
Through transformation of a primary field,
the random process $\scr{G}_{t}$ gives a one-parameter family of germs $\{g_{t}(z)\}_{t\ge 0}$ of holomorphic functions at infinity.
At the same time, each $\scr{G}_{t}$ acts on a highest weight representation of the Virasoro algebra.
In particular, $\scr{G}_{t}\ket{h, c}$ for a certain highest weight vector $\ket{h,c}$ of conformal weight $h$ and central charge $c$ is a random process on
the corresponding highest weight Verma module.
The SLE/CFT correspondence in the sense of Bauer and Bernard states that, if the family of germs $\{g_{t}(z)\}_{t\ge 0}$ satisfies Eq. (\ref{eq:sle}), then
the random process $\scr{G}_{t}\ket{h,c}$ is a martingale for a certain choice of the highest weight because of the existence of a null vector of conformal weight $h+2$
in the irreducible representation $L(h,c)$.
The group theoretical formulation of SLE/CFT correspondence has been convenient for generalizing SLE
by constructing stochastic differential equations associated with more general null vectors in highest weight representations of
the Virasoro algebra \cite{LesageRasmussen2004}, ones for the $\mcal{N}=1$ superconformal algebra \cite{Rasmussen2004a,NagiRasmussen2005},
and ones in logarithmic representations of the Virasoro algebra \cite{Rasmussen2004b, Moghimi-AraghiRajabpourRouhani2004}.

Aside from the group theoretical formulation, there is a {\it correlation function} formulation,
in which one directly investigates martingale conditions on correlation functions to discover the relation between SLE and CFT.
By the correlation function formulation, SLE/CFT correspondence for multiple SLEs has also been established in \cite{BauerBernardKytola2005}
and for further variants of SLE in \cite{Kytola2007}.

CFTs to be considered next to the Virasoro algebra include Wess--Zumino--Witten (WZW) theories associated with
the representation theory of affine Lie algebras.
SLEs corresponding to WZW theories have been considered group theoretically in \cite{Rasmussen2007} for the $\mfrak{sl}_{2}$ case
and in the correlation function formulation in \cite{BettelheimGruzbergLudwigWiegmann2005, AlekseevBytskoIzyurov2011},
which is extended to multiple SLEs in \cite{Sakai2013} and to coset WZW theories in \cite{Nazarov2012,Fukusumi2017}.
The purpose of this paper is to better understand the construction of SLE-type processes in \cite{Rasmussen2007}
to extend it to general WZW theories and to generalize SLE-type growth processes corresponding to WZW theories.
To this purpose, we imitate the construction in \cite{BauerBernard2003a} by considering a random process
on the negative part of the affine Virasoro group, instead of the Virasoro group in \cite{BauerBernard2003a}.
This enables us to obtain an SLE-type growth process associated with a null vector of a certain type in the state space of a WZW theory,
which reduces to the stochastic process constructed in \cite{AlekseevBytskoIzyurov2011} in a particular case.
We also propose new SLE-type growth processes associated with null vectors in the basic representations of $\what{\mfrak{sl}}_{2}$ and $\what{\mfrak{sl}}_{3}$
and conjecture that a similar construction is possible for all $\what{\mfrak{sl}}_{n}$.

Besides the construction of new stochastic processes, a possible application of our formulation is
to establish the module structure of the space of SLE martingales over an affine Lie algebra.
Indeed, consideration on a random process on the Virasoro group was essential 
to define the Virasoro module structure on the space of SLE martingales in \cite{BauerBernard2004a, Kytola2007}.
Similarly, a random process on an affine Virasoro group helps in
finding the affine Lie algebra module structure on the space of SLE martingales,
which we believe is one of completed formulations of SLE/CFT correspondence.
Studies in this direction will be reported elsewhere \cite{SK2018a}.

This paper is organized as follows.
In Sect. \ref{sect:wzw_theory}, we give an introduction to the representation theory of affine Lie algebras
in a somewhat vertex operator algebraic way.
In Sect. \ref{sect:Ito_process}, we formally introduce a random process on a Lie group,
which is a fundamental object in the group theoretical formulation of SLE/CFT correspondence,
and explain our strategy for constructing SLE-type growth processes.
In Sect. \ref{sect:sde}, we derive stochastic differential equations from a given random process on an infinite-dimensional Lie group,
and in Sect. \ref{sect:null_vectors}, we give examples of null vectors, which lead to SLE-type growth processes.

\section{Affine Lie algebras and WZW theories}
\label{sect:wzw_theory}
Let $\mfrak{g}$ be a finite-dimensional simple Lie algebra.
We denote a symmetric invariant bilinear form on $\mfrak{g}$ by $(\cdot|\cdot):\mfrak{g}\times\mfrak{g}\to\mbb{C}$.
The affine Lie algebra of $\mfrak{g}$ is a Lie algebra
\begin{equation}
	\what{\mfrak{g}}=\mfrak{g}\otimes\mbb{C}[t,t^{-1}]\oplus\mbb{C}K
\end{equation}
with Lie brackets being defined by
\begin{align}
	\label{eq:def_of_affine}
	[X\otimes t^{m},Y\otimes t^{n}]&=[X,Y]\otimes t^{m+n}+m(X|Y)\delta_{m+n,0}K, && X, Y\in \mfrak{g},\\
	[K,\what{\mfrak{g}}]&=\{0\}.
\end{align}
As usual, we write $X_{m}$ for $X\otimes t^{m}$.

The state space of the WZW theory associated with $\mfrak{g}$ is a representation of the affine Lie algebra $\what{\mfrak{g}}$.
Let us fix a Cartan subalgebra $\mfrak{h}$ of $\mfrak{g}$,
and, correspondingly, let $\Delta$ and $\Pi^{\vee}=\{\alpha^{\vee}_{1},\cdots,\alpha^{\vee}_{\ell}\}$ be the set of roots and the set of simple coroots, respectively.
The weight lattice is denoted by $P=\bigoplus_{i=1}^{\ell}\mbb{Z}\Lambda_{i}$, where $\Lambda_{i}$ are fundamental weights
defined by $\braket{\Lambda_{i},\alpha^{\vee}_{j}}=\delta_{ij}$.
The set of dominant weights is $P_{+}=\{\Lambda\in P|\braket{\Lambda,\alpha^{\vee}_{i}}\in\mbb{Z}_{\ge 0}\}$.
We denote by $L(\Lambda)$ the finite-dimensional irreducible representation of $\mfrak{g}$ of highest weight $\Lambda\in P_{+}$.
Given a finite-dimensional representation of $\mfrak{g}$, we can construct a representation of the affine Lie algebra $\what{\mfrak{g}}$
by means of parabolic induction.
Let $M$ be a representation of $\mfrak{g}$.
Then the representation $M_{k}$ of a parabolic subalgebra $\what{\mfrak{p}}:=\mfrak{g}\otimes \mbb{C}[t]\oplus\mbb{C}K$ of the affine Lie algebra
is $M$ as a vector space, on which $\mfrak{g}\otimes t^{0}$ acts naturally, $\mfrak{g}\otimes t\mbb{C}[t]$ acts trivially,
and $K$ acts as the multiplication by $k\in \mbb{C}$.
Here the scalar $k$ is called the level.
Then a representation $\what{M}_{k}$ of the affine Lie algebra is obtained by induction:
\begin{equation}
	\what{M}_{k}:=\mrm{Ind}_{\what{\mfrak{p}}}^{\what{\mfrak{g}}}M_{k}=U(\what{\mfrak{g}})\otimes_{U(\what{\mfrak{p}})}M_{k}.
\end{equation}
By the Poincar\'{e}--Birkhoff--Witt theorem, it is isomorphic to $U(\mfrak{g}\otimes t^{-1}\mbb{C}[t^{-1}])\otimes_{\mbb{C}}M_{k}$ as a vector space.
For a finite-dimensional irreducible representation $L(\Lambda)$ of $\mfrak{g}$, $\what{L(\Lambda)}_{k}$, called a Weyl module,
may be reducible for specific levels.
We denote by $L_{\mfrak{g},k}(\Lambda)$ its irreducible quotient.

A WZW theory is specified by a pair $(\mfrak{g},k)$ of a finite-dimensional simple Lie algebra $\mfrak{g}$ and a positive integer level $k\in\mbb{Z}_{>0}$.
The state space of the WZW theory of $(\mfrak{g},k)$ is identified with
\begin{equation}
	V_{\mfrak{g},k}=\bigoplus_{\Lambda\in P_{+}^{k}}L_{\mfrak{g},k}(\Lambda),
\end{equation}
where $P_{+}^{k}=\{\Lambda\in P_{+}|\braket{\Lambda,\theta}\le k\}$ is the set of dominant weights of level $k$,
where $\theta$ is the highest root of $\mfrak{g}$ \cite{FrenkelZhu1992}.

To explain the Sugawara construction, we introduce the notion of field operators.
For $X\in \mfrak{g}$, the corresponding field $X(z)$ is an $\mrm{End}(V_{\mfrak{g},k})$-valued formal power series in a formal variable $z$:
\begin{equation}
	X(z)=\sum_{n\in\mbb{Z}}X_{n}z^{-n-1},
\end{equation}
where the coefficients $X_{n}\in\mrm{End}(V_{\mfrak{g},k})$ are thought of as the action of $X_{n}\in\what{\mfrak{g}}$ on $V_{\mfrak{g},k}$.
From the definition of the affine Lie algebra, Eq. (\ref{eq:def_of_affine}), two fields satisfy the commutation relation
\begin{equation}
	[X(z),Y(w)]=[X,Y](w)\delta(z-w)+k(X|Y)\del_{w}\delta(z-w),
\end{equation}
where $\delta(z-w)=\sum_{n\in\mbb{Z}}z^{-n-1}w^{n}$ is the formal delta distribution.
This commutation relation is often expressed equivalently in the form of operator product expansion (OPE):
\begin{equation}
	X(z)Y(w)\sim \frac{[X,Y](w)}{z-w}+\frac{k(X|Y)}{(z-w)^{2}}.
\end{equation}
For two fields $X(z)$ and $Y(z)$, their naive product $X(z)Y(z)$ is not well-defined
because of the singularity at $z=w$ in the above OPE.
Instead, we define the normal ordered product of two fields.
For a field $X(z)$, its positive and negative power parts $X(z)_{\pm}$ are defined by
\begin{equation}
	X(z)_{+}=\sum_{n=-\infty}^{-1}X_{n}z^{-n-1},\ \ \ X(z)_{-}=\sum_{n=0}^{\infty}X_{n}z^{-n-1}.
\end{equation}
Then the normal ordered product of $X(z)$ and $Y(z)$ is defined by
\begin{equation}
	\no{X(z)Y(z)}=X(z)_{+}Y(z)+Y(z)X(z)_{-}.
\end{equation}

Let $\{X_{i}\}_{i=1}^{\dim\mfrak{g}}$ be an orthonormal basis of $\mfrak{g}$ with respect to the symmetric invariant bilinear form $(\cdot|\cdot)$.
We also assume that the bilinear form is normalized so that $(\theta|\theta)=2$.
We consider the field
\begin{equation}
	L(z)=\frac{1}{2(k+h^{\vee})}\sum_{i=1}^{\dim\mfrak{g}}\no{X_{i}(z)^{2}},
\end{equation}
where $h^{\vee}$ is the dual Coxeter number of $\mfrak{g}$.
Then it satisfies
\begin{equation}
	[L(z),L(w)]=\frac{c_{k}}{2}\del_{w}^{(3)}\delta(z-w)+2L(w)\del_{w}\delta(z-w)+\del L(w)\delta(z-w)
\end{equation}
with $c_{k}=k\dim \mfrak{g}/(k+h^{\vee})$,
which implies that the coefficients of the expansion $L(z)=\sum_{n\in\mbb{Z}}L_{n}z^{-n-2}$ give
a representation of the Virasoro algebra of the central charge $c_{k}$ on $V_{\mfrak{g},k}$.
Moreover, each direct summand $L_{\mfrak{g},k}(\Lambda)$ of $V_{\mfrak{g},k}$, and thus $V_{\mfrak{g},k}$ itself,
decomposes into a direct sum of eigenspaces of $L_{0}$ so that
$L_{\mfrak{g},k}(\Lambda)=\bigoplus_{n\in\mbb{Z}_{\ge 0}}L_{\mfrak{g},k}(\Lambda)_{h_{\Lambda}+n}$,
where
\begin{equation}
	L_{\mfrak{g},k}(\Lambda)_{h}:=\{v\in L_{\mfrak{g},k}(\Lambda)|L_{0}v=hv\}
\end{equation}
is a finite-dimensional eigenspace of $L_{0}$ and the constant $h_{\Lambda}$ is given by
\begin{equation}
	h_{\Lambda}=\frac{(\Lambda|\Lambda+2\rho)}{2(k+h^{\vee})}
\end{equation}
with $\rho=\sum_{i=1}^{\ell}\Lambda_{i}\in \mfrak{h}^{\ast}$ being the Weyl vector.
From direct computation, we find that each $X(z)$ for $X\in\mfrak{g}$ is a primary field with respect to $L(z)$:
\begin{equation}
	\label{eq:primary_aff_field}
	[L(z),X(w)]=X(w)\del_{w}\delta(z-w)+\del X(w)\delta(z-w),
\end{equation}
which, in particular, implies that the mode $X_{-n}$ raises the eigenvalue of $L_{0}$ by $n$.
We also remark that the {\it top space} $L_{\mfrak{g},k}(\Lambda)_{h_{\Lambda}}$ is isomorphic to
the irreducible representation $L(\Lambda)$ of $\mfrak{g}$.

In the rest of this section, we give a vertex operator algebra (VOA) structure on $V_{\mfrak{g},k}$,
which is convenient in the following computation.
To be precise, only the space $L_{\mfrak{g},k}:=L_{\mfrak{g},k}(0)\subset V_{\mfrak{g},k}$ carries a VOA structure,
and the other direct summands are regarded as modules over this VOA.
In this paper, however, we sacrifice the preciseness and introduce the standard computational tools in the theory of VOAs,
without distinguishing {\it intertwining operators} from the state-field correspondence map.
For precise formulation, see \cite{Kac1998, FrenkelBen-Zvi2004, FrenkelZhu1992}.
The fundamental object in the theory of VOAs is the state-field correspondence map
\begin{equation}
	\mcal{Y}(-,z):V_{\mfrak{g},k}\to\mrm{End}(V_{\mfrak{g},k})\{z\}=\left\{\sum_{a\in\mbb{C}}T_{a}z^{a}\bigg|T_{a}\in\mrm{End}(V_{\mfrak{g},k})\right\},
\end{equation}
which satisfies
\begin{equation}
	\mcal{Y}(v,z)\ket{0}|_{z=0}=v
\end{equation}
for an arbitrary $v\in V_{\mfrak{g},k}$.
Here $\ket{0}\in (L_{\mfrak{g},k})_{0}=\mbb{C}\ket{0}$ is a fixed vector called the vacuum.
Another significant property of $\mcal{Y}(-,z)$ is that its value is a field; {\it i.e.},
for arbitrary $v, w\in V_{\mfrak{g},k}$, there are finitely many weights $h_{i}\in\mbb{R}$ ($i=1,\dots,n$) such that
\begin{equation}
	\mcal{Y}(v,z)w\in \bigoplus_{i=1}^{n}V_{\mfrak{g},k}((z))z^{h_{i}}.
\end{equation}
In this notation, the fields $X(z)$ for $X\in\mfrak{g}$ and $L(z)$ are described as
\begin{align}
	X(z)&=\mcal{Y}(X_{-1}\ket{0},z), \\
	L(z)&=\mcal{Y}(L_{-2}\ket{0},z).
\end{align}
Other important fields include one corresponding to $v\in L_{\mfrak{g},k}(\Lambda)_{h_{\Lambda}}$
in the top space of a direct summand of $V_{\mfrak{g},k}$.
This kind of field is often called a {\it primary} field and is characterized by the following properties:
\begin{align}
	\label{eq:primary_cond1}
	[L(z),\mcal{Y}(v,w)]&=h_{\Lambda}\mcal{Y}(v,w)\del_{w}\delta(z-w)+\del\mcal{Y}(v,w)\delta(z-w), \\
	\label{eq:primary_cond2}
	[X(z),\mcal{Y}(v,w)]&=\mcal{Y}(Xv,w)\delta(z-w).
\end{align}
Notice that, for each $\Lambda\in P_{+}^{k}$, there are $\dim L(\Lambda)$ linearly independent primary fields.
The multiplet of them is regarded as an $L(\Lambda)^{\ast}\otimes \mrm{End}(V_{\mfrak{g},k})$-valued field.
\section{Ito process on a Lie group}
\label{sect:Ito_process}
In this section, we formally introduce an Ito process on a Lie group, which is needed in the following.
For a precise definition, see \cite{Applebaum2014}.
Let $\mfrak{g}$ be a finite-dimensional complex Lie algebra that is faithfully represented on a finite-dimensional vector space $V$.
There is a Lie group $G$ realized as a Lie subgroup of $\mrm{GL}(V)$ such that
the Lie algebra of $G$ is $\mfrak{g}$.
We take independent Brownian motions $B_{t}^{(i)}$, $i=1,\dots,\dim\mfrak{g}$, 
and let $\kappa_{i}>0$ be the variances of $B_{t}^{(i)}$; {\it i.e.}, we have
\begin{equation}
	\label{eq:Brownian_motion_cov}
	dB_{t}^{(i)}\cdot dB_{t}^{(j)}=\delta_{i,j}\kappa_{i} dt.
\end{equation}
An Ito process $\scr{G}_{t}$ on the Lie group $G$ is described by a stochastic differential equation
\begin{equation}
	\label{eq:increment_fin_dim}
	\scr{G}_{t}^{-1}d\scr{G}_{t}=\left(\sum_{i}\mu_{i}X_{i}+\frac{1}{2}\sum_{i}\kappa_{i}\sigma_{i}^{2}X_{i}^{2}\right)dt+\sum_{i}\sigma_{i}X_{i}dB_{t}^{(i)},
\end{equation}
where $\{X_{i}\}_{i=1}^{\dim\mfrak{g}}$ is a certain basis of $\mfrak{g}$, and $\mu_{i}$ and $\sigma_{i}$ are random processes satisfying some finiteness conditions.
Note that, in this description, terms $X_{i}^{2}$ do not lie in the Lie algebra $\mfrak{g}$ but are understood as elements in $\mrm{End}(V)$.

Although we assumed that the Lie group $G$ is finite dimensional above,
an Ito process on an infinite-dimensional Lie group is significant in the context of SLE/CFT correspondence.
We assume that the expression in Eq. (\ref{eq:increment_fin_dim}) makes sense on some representations
even if the Lie group is infinite dimensional in the following discussion.

To construct an SLE-type growth process corresponding to a WZW theory,
we consider an Ito process on the subgroup $\what{G}_{\mrm{Vir}}^{<0}$ of the affine Virasoro group.
It is formally generated by exponentials of the negative part $\what{\mfrak{g}}^{<0}_{\mrm{vir}}$ of the affine Virasoro algebra,
which is defined by $\what{\mfrak{g}}^{<0}_{\mrm{vir}}=\mfrak{g}\otimes\mbb{C}[t^{-1}]t^{-1}\oplus\bigoplus_{n\in\mbb{Z}_{<0}}\mbb{C}L_{n}$
with Lie brackets
\begin{align}
	[X\otimes t^{m},Y\otimes t^{n}]&=[X,Y]\otimes t^{m+n}, && X, Y \in\mfrak{g}, \\
	[L_{m},L_{n}]&=(m-n)L_{m+n}, &&\\
	[L_{m},X_{n}]&=-nX_{m+n},&&
\end{align}
for $m,n\in\mbb{Z}_{<0}$.
The commutation relation in Eq. (\ref{eq:primary_aff_field}) suggests that
the state space $V_{\mfrak{g},k}$ is indeed a representation of this Lie algebra.

We assume that an Ito process $\scr{G}_{t}$ on $\what{G}_{\mrm{Vir}}^{<0}$ depends on finitely many mutually independent Brownian motions $\{B_{t}^{(i)}\}_{i\in I_{B}}$
labeled by a certain finite set $I_{B}$
and that the variance of each $B_{t}^{(i)}$ is $\kappa_{i}>0$.
Then, analogously to the expression in Eq. (\ref{eq:increment_fin_dim}), the increment of $\scr{G}_{t}$ can be written in the form
\begin{equation}
	\scr{G}_{t}^{-1}d\scr{G}_{t}=\left(\mu+\frac{1}{2}\sum_{i\in I_{B}}\kappa_{i}\sigma_{i}^{2}\right)dt+\sum_{i\in I_{B}}\sigma_{i}dB_{t}^{(i)},
\end{equation}
where $\mu$ and $\sigma_{i}$ for $i\in I_{B}$ are $\what{\mfrak{g}}_{\mrm{vir}}^{<0}$-valued random processes.
We also impose the initial condition so that $\scr{G}_{0}$ is the unit element in $\what{G}_{\mrm{Vir}}^{<0}$.

As is explained in the previous section,
for a vector $v\in (L_{\mfrak{g},k}(\Lambda))_{h_{\Lambda}}$ in the top space of $L_{\mfrak{g},k}(\Lambda)$,
the corresponding field $\mcal{Y}(v,z)$ satisfies the primary field conditions given by Eqs. (\ref{eq:primary_cond1}) and (\ref{eq:primary_cond2}).
This implies that the adjoint action of $\scr{G}\in \what{G}_{\mrm{Vir}}^{<0}$ transforms the primary field $\mcal{Y}(v,z)$ so that
\begin{equation}
	\scr{G}^{-1}\mcal{Y}(v,z)\scr{G}=(\pr{f}(z))^{h_{\Lambda}}\mcal{Y}(\Theta(z)v,f(z)).
\end{equation}
Here $f(z)\in \mbb{C}[[z^{-1}]]z$ is a formal Laurent series in $z^{-1}$ with a nonzero coefficient of $z$, 
and $\Theta(z)\in G[[z^{-1}]]$ is a $\mbb{C}[[z^{-1}]]$-valued point in the finite-dimensional Lie group $G$ such that $\mrm{Lie}(G)=\mfrak{g}$,
which naturally defines a homomorphism $L(\Lambda)\to L(\Lambda)[[z^{-1}]]$.
We have to comment that $\mcal{Y}(\Theta(z)v,f(z))$ is not a field,
thus we make $\mcal{Y}(\Theta(z)v,f(z))$ meaningful in the following way.
Let $V_{\mfrak{g},k}=\bigoplus_{h\in \mcal{P}}(V_{\mfrak{g},k})_{h}$ be the eigenspace decomposition with respect to $L_{0}$,
where $\mcal{P}=\bigcup_{\Lambda\in P_{+}^{k}}(\mbb{Z}_{\ge 0}+h_{\Lambda})$ is the set of eigenvalues.
Then the restricted dual space of $V_{\mfrak{g},k}$ is defined by $\dg{V}_{\mfrak{g},k}=\bigoplus_{h\in\mcal{P}}(V_{\mfrak{g},k})_{h}^{\ast}$.
Although $\mcal{Y}(\Theta(z)v,f(z))$ is not a field, for a vector $u^{\ast}\in \dg{V}_{\mfrak{g},k}$,
there are finitely many weights $h_{i}\in\mbb{R}$ ($i=1,\dots,n$),
such that $\bra{u^{\ast}}\mcal{Y}(\Theta(z)v,f(z))\in \dg{V}_{\mfrak{g},k}((z^{-1}))z^{h_{i}}$.
We understand $\mcal{Y}(\Theta(z)v,f(z))$ as such an object in the following.

Replacing $\scr{G}$ by the Ito process $\scr{G}_{t}$ on $\what{G}_{\mrm{Vir}}^{<0}$, we obtain
\begin{equation}
	\label{eq:random_transformation_primary}
	\scr{G}_{t}^{-1}\mcal{Y}(v,z)\scr{G}_{t}=(\pr{f}_{t}(z))^{h_{\Lambda}}\mcal{Y}(\Theta_{t}(z)v,f_{t}(z)),
\end{equation}
where $f_{t}(z)$ is a $\mbb{C}[[z^{-1}]]z$-valued Ito process and $\Theta_{t}(z)$ is a $G[[z^{-1}]]$-valued one.
Namely, an Ito process $\scr{G}_{t}$ on $\what{G}_{\mrm{Vir}}^{<0}$ induces a pair of, in general, 
interacting random processes $(f_{t}(z), \Theta_{t}(z))$ on $\mbb{C}[[z^{-1}]]z$ and $G[[z^{-1}]]$.
We suppose that the increment of $f_{t}(z)$ is written as
\begin{equation}
	df_{t}(z)=\overline{f}_{t}(z)dt+\sum_{i\in I_{B}}f_{t}^{i}(z)dB_{t}^{(i)}
\end{equation}
with $\overline{f}_{t}(z)$ and $f_{t}^{i}(z)$ being some $\mbb{C}[[z^{-1}]]z$-valued random processes.
To describe the increment of $\Theta_{t}(z)$, we take a basis $\{X_{r}\}_{r\in I_{\mfrak{g}}}$ of $\mfrak{g}$ with the index set $I_{\mfrak{g}}$.
Then we can express the increment of $\Theta_{t}(z)$ as
\begin{equation}
	d\Theta_{t}(z)\Theta_{t}(z)^{-1}=\sum_{r\in I_{\mfrak{g}}}d\theta_{r,t}(z)X_{r}+\frac{1}{2}\sum_{r,s\in I_{\mfrak{g}}}d\theta_{r,t}(z)d\theta_{s,t}(z)X_{r}X_{s}.
\end{equation}
We also suppose that the increment of each $\theta_{r,t}(z)$ is written as
\begin{equation}
	d\theta_{r,t}(z)=\overline{\theta}_{r,t}(z)dt+\sum_{i\in I_{B}}\theta_{r,t}^{i}(z)dB_{t}^{(i)},
\end{equation}
with $\overline{\theta}_{r,t}(z)$ and $\theta_{r,t}^{i}(z)$ being $\mbb{C}[[z^{-1}]]$-valued random processes.

Our strategy is to compare the increments of both sides of Eq. (\ref{eq:random_transformation_primary})
and to establish stochastic differential equations on $f_{t}(z)$ and $\theta_{r,t}(z)$ for a given Ito process $\scr{G}_{t}$
on $\what{G}_{\mrm{Vir}}^{<0}$.

Notice that, from $d(\scr{G}_{t}^{-1}\scr{G}_{t})=0$, we have
\begin{equation}
	(d\scr{G}_{t}^{-1})\scr{G}_{t}=\left(-\mu+\frac{1}{2}\sum_{i\in I_{B}}\kappa_{i}\sigma_{i}^{2}\right)dt-\sum_{i\in I_{B}}\sigma_{i}dB_{t}^{(i)}.
\end{equation}
Using this formula, we find that the increment of the left-hand side of Eq. (\ref{eq:random_transformation_primary}) yields
\begin{align}
	\label{eq:increment_lhs}
	&d(\scr{G}_{t}^{-1}\mcal{Y}(v,z)\scr{G}_{t}) \\
	&=(\pr{f}_{t}(z))^{h_{\Lambda}}\left(-[\mu,\mcal{Y}(\Theta_{t}(z)v,f_{t}(z))]+\frac{1}{2}\sum_{i\in I_{B}}\kappa_{i}[\sigma_{i},[\sigma_{i},\mcal{Y}(\Theta_{t}(z)v,f_{t}(z))]]\right)dt \notag \\
	&\hspace{11pt}-(\pr{f}_{t}(z))^{h_{\Lambda}}\sum_{i\in I_{B}}[\sigma_{i},\mcal{Y}(\Theta_{t}(z)v,f_{t}(z))]dB_{t}^{(i)}. \notag
\end{align}
The increment of the right-hand side of Eq. (\ref{eq:random_transformation_primary}) is computed by standard Ito calculus as
\begin{align}
	\label{eq:increment_rhs}
	&d\left((\pr{f}_{t}(z))^{h_{\Lambda}}\mcal{Y}(\Theta_{t}(z)v,f(z))\right) \\
	&=(\pr{f}_{t}(z))^{h_{\Lambda}}\Bigg\{\bigg(h_{\Lambda}\frac{\overline{f}^{\prime}_{t}(z)}{\pr{f}_{t}(z)}+\frac{1}{2}h_{\Lambda}(h_{\Lambda}-1)\sum_{i\in I_{B}}\kappa_{i}\left(\frac{f_{t}^{i\prime}(z)}{\pr{f}_{t}(z)}\right)^{2} \notag \\
	&\hspace{70pt}+\overline{f}_{t}(z)\frac{\del}{\del f_{t}(z)}+h_{\Lambda}\sum_{i\in I_{B}} \kappa_{i}\frac{f_{t}^{i\prime}(z)}{\pr{f}_{t}(z)}f_{t}^{i}(z)\frac{\del}{\del f_{t}(z)} \notag \\
	&\hspace{70pt}+\frac{1}{2}\sum_{i\in I_{B}}\kappa_{i}f_{t}^{i}(z)^{2}\frac{\del^{2}}{\del f_{t}(z)^{2}}\bigg)\mcal{Y}(\Theta_{t}(z)v,f_{t}(z)) \notag \\
	&\hspace{70pt}+\sum_{r\in I_{\mfrak{g}}}\bigg(\overline{\theta}_{r,t}(z)+h_{\Lambda}\sum_{i\in I_{B}}\kappa_{i}\frac{f_{t}^{i\prime}(z)}{\pr{f}_{t}(z)}\theta_{r,t}^{i}(z) \notag \\
	&\hspace{105pt}+\sum_{i\in I_{B}}\kappa_{i}\theta_{r,t}^{i}(z)f_{t}^{i}(z)\frac{\del}{\del f_{t}(z)}\bigg)\mcal{Y}(X_{r}\Theta_{t}(z)v,f_{t}(z)) \notag \\
	&\hspace{70pt}+\frac{1}{2}\sum_{r,s \in I_{\mfrak{g}}}\sum_{i\in I_{B}}\kappa_{i}\theta_{r,t}^{i}(z)\theta_{s,t}^{i}(z)\mcal{Y}(X_{r}X_{s}\Theta_{t}(z)v,f_{t}(z))\Bigg\}dt \notag \\
	&\hspace{11pt}+(\pr{f}_{t}(z))^{h_{\Lambda}}\sum_{i\in I_{B}}\Bigg\{\bigg(h_{\Lambda}\frac{f_{t}^{i\prime}(z)}{\pr{f}_{t}(z)}+f_{t}^{i}(z)\frac{\del}{\del f_{t}(z)}\bigg)\mcal{Y}(\Theta_{t}(z)v,f_{t}(z)) \notag \\
	&\hspace{90pt}+\sum_{r\in I_{\mfrak{g}}}\theta_{r,t}^{i}(z)\mcal{Y}(X_{r}\Theta_{t}(z)v,f_{t}(z))\Bigg\}dB_{t}^{(i)}. \notag
\end{align}
Comparing the coefficients of each $dB_{t}^{(i)}$ in Eqs. (\ref{eq:increment_lhs}) and (\ref{eq:increment_rhs}), we find
\begin{align}
	[\sigma_{i},\mcal{Y}(\Theta_{t}(z)v,f_{t}(z))]
	&=-\left(h_{\Lambda}\frac{f_{t}^{i\prime}(z)}{\pr{f}_{t}(z)}+f_{t}^{i}(z)\frac{\del}{\del f_{t}(z)}\right)\mcal{Y}(\Theta_{t}(z)v,f_{t}(z))  \\
	&\hspace{11pt}-\sum_{r\in I_{\mfrak{g}}}\theta_{r,t}^{i}(z)\mcal{Y}(X_{r}\Theta_{t}(z)v,f_{t}(z)). \notag
\end{align}
The primary field conditions, Eqs. (\ref{eq:primary_cond1}) and (\ref{eq:primary_cond2}), help us
determine $f_{t}^{i}(z)$ and $\theta_{r,t}^{i}(z)$ for a given $\sigma_{i}$.
For example, if $\sigma_{i}=L_{n}$, we have
\begin{align}
	f_{t}^{i}(z)&=-(f_{t}(z))^{n+1},\\
	\theta_{r,t}^{i}(z)&=0,\ \ \ \mbox{for all }r\in I_{\mfrak{g}}.
\end{align}
Also, if $\sigma_{i}=-(X_{r})_{n}$, we have
\begin{align}
	f_{t}^{i}(z)&=0, \\
	\theta_{r,t}^{i}(z)&=(f_{t}(z))^{n}, \\
	\theta_{s,t}^{i}(z)&=0,\ \ \ \mbox{for }s\neq r.
\end{align}
These give part of stochastic differential equations on $f_{t}(z)$ and $\theta_{r,t}(z)$.
Comparing the coefficients of $dt$ in Eqs. (\ref{eq:increment_lhs}) and (\ref{eq:increment_rhs}),
we can fully determine these stochastic differential equations, in principle.
In the next section, we complete this program for a certain choice of $\mu$, $\sigma_{i}$.

\section{Derivations of stochastic differential equations}
\label{sect:sde}
Let $\{X_{r}\}_{r\in I_{\mfrak{g}}}$ be an orthonormal basis of $\mfrak{g}$
with respect to the symmetric invariant bilinear form $(\cdot|\cdot)$.
We prepare mutually independent Brownian motions $B_{t}^{(i)}$ labeled by $i\in I_{B}:=\{0\}\cup I_{\mfrak{g}}$, identifying $I_{\mfrak{g}}$ with a subset of $I_{B}$,
and let $\kappa_{i}>0$ for $i \in I_{B}$ be the variance of $B_{t}^{(i)}$.

We also assume that $\mu=-2L_{-2n}$, $\sigma_{0}=L_{-n}$, and $\sigma_{r}=-(X_{r})_{-n}$ for $i\in I_{\mfrak{g}}$ with $n\in \mbb{Z}_{>0}$.
Then the Ito process $\scr{G}_{t}$ on $\what{G}_{\mrm{Vir}}^{<0}$ is described by
\begin{align}
	\label{eq:stochastic_dif_specific}
	\scr{G}_{t}^{-1}d\scr{G}_{t}
	&=\left[-2L_{-2n}+\frac{1}{2}\kappa_{0} L_{-n}^{2}+\frac{1}{2}\sum_{r\in I_{\mfrak{g}}}\kappa_{r}(X_{r})_{-n}^{2}\right]dt  \\
	&\hspace{11pt}+L_{-n}dB_{t}^{(0)}-\sum_{r\in I_{\mfrak{g}}}(X_{r})_{-n}dB_{t}^{(r)}. \notag
\end{align}
We remark that, even if $n=1$, the above Ito process $\scr{G}_{t}$ is different from the one considered in \cite{Rasmussen2007},
which does not contain Brownian motion along $L_{-1}$.
The last discussion in the previous section suggests that
\begin{align}
	&f_{t}^{0}(z)=-(f_{t}(z))^{-n+1},\ \ \theta_{r,t}^{0}(z)=0,\ \ r\in I_{\mfrak{g}}, \\
	&f_{t}^{r}(z)=0,\ \ \theta_{r,t}^{r}(z)=(f_{t}(z))^{-n},\ \ \theta_{s,t}^{r}(z)=0, \ \ r\in I_{\mfrak{g}},\ s \neq r.
\end{align}
These results simplify Eq. (\ref{eq:increment_rhs}) to
\begin{align}
	&d\left((\pr{f}_{t}(z))^{h_{\Lambda}}\mcal{Y}(\Theta_{t}(z)v,f(z))\right) \\
	&=(\pr{f}_{t}(z))^{h_{\Lambda}}\Bigg\{\bigg(h_{\Lambda}\frac{\overline{f}^{\prime}_{t}(z)}{\pr{f}_{t}(z)}+\frac{1}{2}h_{\Lambda}(h_{\Lambda}-1)(n-1)^{2}\kappa_{0}(f_{t}(z))^{-2n} \notag \\
	&\hspace{70pt}+\left(\overline{f}_{t}(z)+h_{\Lambda}\kappa_{0}(-n+1)(f_{t}(z))^{-2n+1}\right)\frac{\del}{\del f_{t}(z)} \notag \\
	&\hspace{70pt}+\frac{1}{2}\kappa_{0}(f_{t}(z))^{-2n+2}\frac{\del^{2}}{\del f_{t}(z)^{2}}\bigg)\mcal{Y}(\Theta_{t}(z)v,f_{t}(z)) \notag \\
	&\hspace{70pt}+\sum_{r\in I_{\mfrak{g}}}\overline{\theta}_{r,t}(z)\mcal{Y}(X_{r}\Theta_{t}(z)v,f_{t}(z)) \notag \\
	&\hspace{70pt}+\frac{1}{2}\sum_{r\in I_{\mfrak{g}}}\kappa_{r}(f_{t}(z))^{-2n}\mcal{Y}(X_{r}^{2}\Theta_{t}(z)v,f_{t}(z))\Bigg\}dt \notag \\
	&\hspace{11pt}+(\mbox{terms proportional to }dB_{t}^{(i)}). \notag
\end{align}
The increment in Eq. (\ref{eq:increment_lhs}) becomes
\begin{align}
	&d(\scr{G}_{t}^{-1}\mcal{Y}(v,z)\scr{G}_{t}) \\
	&=(\pr{f}_{t}(z))^{h_{\Lambda}}\Bigg\{\bigg(2\left(h_{\Lambda}(-2n+1)+\frac{\kappa_{0}}{2}(h_{\Lambda}^{2}(n-1)^{2}+h_{\Lambda}n(n-1))\right)(f_{t}(z))^{-2n} \notag \\
	&\hspace{70pt}+\left(2+\frac{\kappa_{0}}{2}(-n+1)+h_{\Lambda}\kappa_{0}(-n+1)\right)f_{t}(z)^{-2n+1}\frac{\del}{\del f_{t}(z)} \notag \\
	&\hspace{70pt}+\frac{1}{2}\kappa_{0}(f_{t}(z))^{-2n+2}\frac{\del^{2}}{\del f_{t}(z)^{2}}\bigg)\mcal{Y}(\Theta_{t}(z)v,f_{t}(z)) \notag \\
	&\hspace{70pt}+\frac{1}{2}\sum_{r\in I_{\mfrak{g}}}\kappa_{r}(f_{t}(z))^{-2n}\mcal{Y}(X_{r}^{2}\Theta_{t}(z)v,f_{t}(z))\Bigg\}dt \notag \\
	&\hspace{11pt}+(\mbox{terms proportional to }dB_{t}^{(i)}). \notag
\end{align}
Then we compare the coefficients of $dt$ to obtain
\begin{align}
	&\overline{f}_{t}(z)=\left(2+\frac{1}{2}\kappa_{0}(-n+1)\right)(f_{t}(z))^{-2n+1}, \\
	&\overline{\theta}_{r,t}(z)=0,\ \ r\in I_{\mfrak{g}}.
\end{align}
Therefore, the Ito processes $f_{t}(z)$ and $\theta_{k,t}(z)$ satisfy the following stochastic differential equations:
\begin{align}
	\label{eq:growth_process_geom}
	df_{t}(z)&=\left(2-\frac{1}{2}\kappa_{0}(n-1)\right)(f_{t}(z))^{-2n+1}dt-(f_{t}(z))^{-n+1}dB_{t}^{(0)}, \\
	\label{eq:growth_process_alg}
	d\theta_{r,t}(z)&=(f_{t}(z))^{-n}dB_{t}^{(r)},\ \ r\in I_{\mfrak{g}}.
\end{align}
If $n=1$, the differential equations (\ref{eq:growth_process_geom}) and (\ref{eq:growth_process_alg})
reduce to the ones considered in \cite{AlekseevBytskoIzyurov2011}.
Therefore, one can regard the set of equations (\ref{eq:growth_process_geom}) and (\ref{eq:growth_process_alg})
as generalization of the SLE-type growth process obtained in \cite{AlekseevBytskoIzyurov2011}.

If we define an Ito process $g_{t}(z)$ by
\begin{equation}
	g_{t}(z)=f_{t}(z)^{n}+nB_{t}^{(0)},
\end{equation}
we obtain the following set of stochastic differential equations:
\begin{align}
	\label{eq:sle_rescaled}
	dg_{t}(z)&=\frac{2n}{g_{t}(z)-nB_{t}^{(0)}}dt, \\
	d\theta_{r,t}(z)&=\frac{dB_{t}^{(r)}}{g_{t}(z)-nB_{t}^{(0)}},\ \ r\in I_{\mfrak{g}}.
\end{align}
It is remarkable that Eq. (\ref{eq:sle_rescaled}) is nothing but the stochastic Loewner equation (\ref{eq:sle}) with a rescaled time variable,
but we have to comment that here the introduced $g_{t}(z)$ cannot be a uniformizing map in the sense of \cite{BauerBernard2003a} unless $n=1$, while $f_{t}(z)$ can be.
To see this, let us consider the automorphism group of an algebra of formal power series.
Let $\mcal{O}:=\mbb{C}[[w]]=\varprojlim \mbb{C}[w]/(w^{n})$ be a complete $\mbb{C}$-algebra.
A continuous automorphism of $\mcal{O}$ is identified with the image of the topological generator $w$ of $\mcal{O}$.
Under this identification, the group of continuous automorphisms $\mrm{Aut}\mcal{O}$ of $\mcal{O}$ is described as
\begin{equation}
	\mrm{Aut}\mcal{O}\simeq \left\{w\mapsto a_{1}w+a_{2}w^{2}+\cdots| a_{1}\in \mbb{C}^{\times},\ a_{i}\in\mbb{C}\ (i\ge 2)\right\}.
\end{equation}
By setting $w=1/z$, we regard the formal disk $D=\mrm{Spec}\mcal{O}$ as the formal neighborhood at infinity.
Then $f_{t}(z)$ constructed in our formulation defines a continuous automorphism of $\mcal{O}$ identified with
\begin{equation}
	w\mapsto \frac{1}{f(1/w)},
\end{equation}
while $g_{t}(z)$ does not if $n\ge 2$.
Indeed, $1/(g_{t}(1/w))$ begins from $w^{n}$ in $\mcal{O}$, and, in particular, it is not invertible.

\section{Null vectors and martingale conditions}
\label{sect:null_vectors}
For a certain vector $w\in (L_{\mfrak{g},k}(\Lambda))_{h_{\Lambda}}$, $\scr{G}_{t}w$ is an Ito process on the representation $L_{\mfrak{g},k}(\Lambda)$.
It is an important problem when this random process is a martingale.
To explain its importance, let us consider a correlation function or, precisely, a correlation-function-valued random process of the form
\begin{equation}
	\label{eq:correlation_function1}
	\braket{0|\mcal{Y}(v_{1},z_{1})\cdots\mcal{Y}(v_{N},z_{N})\scr{G}_{t}w}.
\end{equation}
Here we take $v_{i}\in (L_{\mfrak{g},k}(\Lambda_{i}))_{h_{\Lambda_{i}}}$ and
we let $\bra{0}$ be the dual of the vacuum vector $\ket{0}\in L_{\mfrak{g},k}$.
From the transformation rule of primary fields, Eqs. (\ref{eq:primary_cond1}) and (\ref{eq:primary_cond2}),
and the translation invariance of the vacuum vector,
the above function is the same as
\begin{equation}
	\label{eq:correlation_function2}
	\prod_{i=1}^{N}(\pr{f}_{t}(z_{i}))^{h_{\Lambda_{i}}}\braket{0|\mcal{Y}(\Theta_{t}(z)v_{1},f_{t}(z_{1}))\cdots\mcal{Y}(\Theta_{t}(z)v_{N},f_{t}(z_{N}))w}
\end{equation}
As is demonstrated in \cite{BauerBernard2003a}, the expectation value of Eq. (\ref{eq:correlation_function2})
has a probability theoretical interpretation.
If $\scr{G}_{t}w$ is a martingale, the expectation value of Eq. (\ref{eq:correlation_function1})
does not depend on time, yielding
\begin{equation}
	\mbb{E}[\braket{0|\mcal{Y}(v_{1},z_{1})\cdots\mcal{Y}(v_{N},z_{N})\scr{G}_{t}w}]
	=\braket{0|\mcal{Y}(v_{1},z_{1})\cdots\mcal{Y}(v_{N},z_{N})w},
\end{equation}
the right-hand side of which is a purely algebraic object computed in CFT.
Therefore, the martingale property of $\scr{G}_{t}$ is significant in bridging probability theory to CFT via SLE/CFT correspondence.

For our $\scr{G}_{t}$ designed in Eq. (\ref{eq:stochastic_dif_specific}),
the random process $\scr{G}_{t}w$ on $L_{\mfrak{g},k}(\Lambda)$ is a martingale if and only if the vector
\begin{equation}
	\label{eq:specific_null_vector}
	\left[-2L_{-2n}+\frac{1}{2}\kappa_{0} L_{-n}^{2}+\frac{1}{2}\sum_{r\in I_{\mfrak{g}}}\kappa_{r}(X_{r})_{-n}^{2}\right]w
\end{equation}
is a null vector in the corresponding Weyl module.
For $n=1$, the conditions for the above vector to be a singular vector have been analyzed in \cite{AlekseevBytskoIzyurov2011}
for the case when $\kappa_{r}$ are the same constant for all $r\in I_{\mfrak{g}}$.

Here, we give examples of null vectors of the above form for $n=2$ in the basic representations of $\what{\mfrak{sl}}_{2}$ and $\what{\mfrak{sl}}_{3}$.
We take an orthonormal basis of $\mfrak{sl}_{2}$ as
\begin{align}
	X_{1}=\frac{1}{\sqrt{2}}H, && X_{2}=\frac{1}{\sqrt{2}}(E+F), && X_{3}=\frac{i}{\sqrt{2}}(E-F).
\end{align}
Here $E,\ H$, and $F$ are the standard basis
\begin{equation}
	E=\left(\begin{array}{cc}
			0	& 1 \\
			0	& 0
		\end{array}\right),\ \ 
	H=\left(\begin{array}{cc}
			1	& 0 \\
			0	& -1
		\end{array}\right),\ \ 
	F=\left(\begin{array}{cc}
			0	& 0 \\
			1	& 0
		\end{array}\right),
\end{equation}
and the symmetric invariant bilinear form is given by $(X|Y)=\mrm{Tr}(XY)$ for $X, Y \in\mfrak{sl}_{2}$.
Then the problem is whether there is a null vector in $L_{\mfrak{sl}_{2},1}$ of the form
\begin{equation}
	\left[-2L_{-4}+\frac{1}{2}\kappa_{0}L_{-2}^{2}+\frac{1}{2}\sum_{r=1}^{3}\kappa_{r}(X_{r})_{-2}^{2}\right]\ket{0}
\end{equation}
with $\kappa_{i}>0$ for $i=0,1,2,3$.

Because the basic representation $L_{\mfrak{sl}_{2},1}$ of $\what{\mfrak{sl}}_{2}$ is isomorphic to
the lattice VOA $V_{Q}$ associated with the root lattice $Q=\mbb{Z}\alpha$ with $(\alpha|\alpha)=2$ of $\mfrak{sl}_{2}$ \cite{FrenkelKac1980},
it is convenient to express the operation of $\what{\mfrak{sl}}_{2}$ in terms of the free Bose field and vertex operators.
The isomorphism is described by the following assignment:
\begin{align}
	E(z)\mapsto \Gamma_{\alpha}(z), && H(z)\mapsto \alpha(z), && F(z)\mapsto \Gamma_{-\alpha}(z).
\end{align}
Here $\alpha(z)$ is the free Bose field, and $\Gamma_{\pm \alpha}(z)$ are vertex operators of charge $\pm \alpha\in Q$.
Then the Virasoro field $L(z)$ is given by the Sugawara construction, which yields
\begin{equation}
	L(z)=\frac{1}{4}\no{\alpha(z)^{2}}.
\end{equation}
In the notion of VOA, the Virasoro field is $\mcal{Y}(L_{-2}\ket{0},z)$ corresponding to
the Virasoro vector $L_{-2}\ket{0}$.
By the translation covariance of fields we have
\begin{equation}
	\mcal{Y}(L_{-4}\ket{0},z)=\frac{1}{2}\del^{2}L(z)=\frac{1}{4}\left(\no{\del^{2}\alpha(z)\alpha(z)}+\no{\del\alpha(z)^{2}}\right).
\end{equation}
By computation of the OPE, we can get
\begin{align}
	&\mcal{Y}(L_{-2}^{2}\ket{0},z)=\no{L(z)^{2}} \\
	&=\frac{1}{4}\no{\del^{2}\alpha(z)\alpha(z)}+\frac{1}{16}\no{\alpha(z)^{4}}, \notag \\
	&\mcal{Y}(E_{-2}F_{-2}\ket{0},z)=\no{\del\Gamma_{\alpha}(z)\del\Gamma_{-\alpha}(z)} \\
	&=\frac{1}{12}\left(\del^{3}\alpha(z)+2\no{\del^{2}\alpha(z)\alpha(z)}+3\no{\del\alpha(z)^{2}}-\no{\alpha(z)^{4}}\right) \notag\\
	&\mcal{Y}(F_{-2}E_{-2}\ket{0},z)=\no{\del\Gamma_{-\alpha}(z)\del\Gamma_{\alpha}(z)} \\
	&=\frac{1}{12}\left(-\del^{3}\alpha(z)+2\no{\del^{2}\alpha(z)\alpha(z)}+3\no{\del\alpha(z)^{2}}-\no{\alpha(z)^{4}}\right) \notag\\
	&\mcal{Y}(H_{-2}^{2}\ket{0},z)=\no{\del\alpha(z)^{2}}.
\end{align}
Combining these, we can see that
\begin{equation}
	\left(-2L_{-4}+\frac{4}{3}L_{-2}^{2}+\frac{1}{2}\sum_{r=1}^{3}(X_{r})_{-2}^{2}\right)\ket{0}=0
\end{equation}
in $L_{\mfrak{sl}_{2},1}$.
Therefore, $\scr{G}_{t}\ket{0}$ corresponding to $n=2$, $\kappa_{0}=\frac{8}{3}$, and $\kappa_{1}=\kappa_{2}=\kappa_{3}=1$
is a martingale.

We can also look for a null vector of the form of Eq. (\ref{eq:specific_null_vector}) in $L_{\mfrak{sl}_{3},1}$ by similar computation.
An orthonormal basis of $\mfrak{sl}_{3}$ is taken as
\begin{align}
	X_{1}&= \frac{1}{\sqrt{2}}H_{1}, & X_{2}&=\frac{1}{\sqrt{6}}(H_{1}+2H_{2}), \\
	X_{3}&=\frac{1}{\sqrt{2}}(E_{1}+F_{1}), & X_{4}&=\frac{i}{\sqrt{2}}(E_{1}-F_{1}), \\
	X_{5}&=\frac{1}{\sqrt{2}}(E_{2}+F_{2}), & X_{6}&=\frac{i}{\sqrt{2}}(E_{2}-F_{2}), \\
	X_{7}&=\frac{1}{\sqrt{2}}([E_{1},E_{2}]+[F_{2},F_{1}]) & X_{8}&=\frac{i}{\sqrt{2}}([E_{1},E_{2}]-[F_{2},F_{1}]).
\end{align}
Here $E_{i}$, $H_{i}$, and $F_{i}$ are defined by
\begin{align}
	E_{1}&=\left(\begin{array}{ccc}
			0  & 1 & 0 \\
			0 & 0 & 0\\
			0 & 0 & 0
		\end{array}\right),&
	H_{1}&=\left(\begin{array}{ccc}
			1  & 0 & 0 \\
			0 & -1 & 0\\
			0 & 0 & 0
		\end{array}\right),&
	F_{1}&=\left(\begin{array}{ccc}
			0  & 0 & 0 \\
			1 & 0 & 0\\
			0 & 0 & 0
		\end{array}\right), \\
	E_{2}&=\left(\begin{array}{ccc}
			0  & 0 & 0 \\
			0 & 0 & 1\\
			0 & 0 & 0
		\end{array}\right),&
	H_{2}&=\left(\begin{array}{ccc}
			0  & 0 & 0 \\
			0 & 1 & 0\\
			0 & 0 & -1
		\end{array}\right),&
	F_{1}&=\left(\begin{array}{ccc}
			0  & 0 & 0 \\
			0 & 0 & 0\\
			0 & 1 & 0
		\end{array}\right),
\end{align}
and the symmetric invariant bilinear form is given by the same formula $(X|Y)=\mrm{Tr}(XY)$ for $X, Y\in\mfrak{sl}_{3}$ as in the case of $\mfrak{sl}_{2}$.
Again, $L_{\mfrak{sl}_{3},1}$ is isomorphic to the lattice VOA associated to the root lattice $Q=\mbb{Z}\alpha_{1}+\mbb{Z}\alpha_{2}$
of type $A_{2}$.
The action of $\what{\mfrak{sl}}_{3}$ is described by the assignment
\begin{align}
	E_{i}(z)\mapsto \Gamma_{\alpha_{i}}(z), && H_{i}(z)\mapsto \alpha_{i}(z), && F_{i}(z)\mapsto \Gamma_{-\alpha_{i}}(z)
\end{align}
for $i=1,2$.
By similar but more complicated computation than in the case of $\what{\mfrak{sl}}_{2}$,
we find that
\begin{equation}
	\left[-2L_{-4}+\frac{6}{5}L_{-2}^{2}+\frac{2}{5}\sum_{r=1}^{8}(X_{r})_{-2}^{2}\right]\ket{0}=0
\end{equation}
in $L_{\mfrak{sl}_{3},1}$.
Therefore, the random process $\scr{G}_{t}\ket{0}$ in $L_{\mfrak{sl}_{3},1}$ is a martingale if $\kappa_{0}=\frac{12}{5}$
and $\kappa_{i}=\frac{4}{5}$ for $i=1,\dots,8$.

From the above observations for $\what{\mfrak{sl}}_{2}$ and $\what{\mfrak{sl}}_{3}$,
we make a conjecture that there is a null vector of the form
\begin{equation}
	\label{eq:null_vector_conjecture}
	\left(-2L_{-4}+\frac{\kappa}{2}L_{-2}^{2}+\frac{\tau}{2}\sum_{r=1}^{n^{2}-1}(X_{r})_{-2}^{2}\right)\ket{0}
\end{equation}
in the basic representation $L_{\mfrak{sl}_{n},1}$ of $\what{\mfrak{sl}}_{n}$.
Here $\{X_{r}\}_{r=1}^{n^{2}-1}$ is an orhonormal basis of $\mfrak{sl}_{n}$ with respect to
the trace form $(X|Y)=\mrm{Tr}(XY)$ for $X, Y\in\mfrak{sl}_{n}$.

The situation is more subtle for a spin representation of level $1$.
It may be impossible to construct a null vector of the form of Eq. (\ref{eq:specific_null_vector}) in $L_{\mfrak{g},1}(\Lambda)$
other than $\Lambda=0$.
Indeed, we have confirmed that there does not exist such a null vector in the spin-$\frac{1}{2}$ representation $L_{\mfrak{sl}_{2},1}(\Lambda_{1})$
of $\what{\mfrak{sl}}_{2}$.

\subsection*{Acknowledgements}
The author is grateful to  K. Sakai and R. Sato for fruitful discussions.
This work was supported by a Grant-in-Aid for JSPS Fellows (Grant No. 17J09658).

\addcontentsline{toc}{chapter}{Bibliography}
\bibliographystyle{alpha}
\bibliography{sle_cft}

\newcommand{\etalchar}[1]{$^{#1}$}
\begin{thebibliography}{CDCH{\etalchar{+}}14}

\bibitem[ABI11]{AlekseevBytskoIzyurov2011}
A.~Alekseev, A.~Bytsko, and K.~Izyurov.
\newblock On {SLE} martingales in boundary {WZW} models.
\newblock {\em Lett. Math. Phys.}, 97:243--261, 2011.

\bibitem[App14]{Applebaum2014}
David Applebaum.
\newblock {\em Probability on Compact Lie Groups}, volume~70 of {\em
  Probability Theory and Stochastic Modeling}.
\newblock Springer, 2014.

\bibitem[BB02]{BauerBernard2002}
M.~Bauer and D.~Bernard.
\newblock {SLE}$_{\kappa}$ growth processes and conformal field theories.
\newblock {\em Phys. Lett. B}, 543:135--138, 2002.

\bibitem[BB03a]{BauerBernard2003a}
M.~Bauer and D.~Bernard.
\newblock Conformal field theories of stochastic {Loewner} evolutions.
\newblock {\em Commun. Math. Phys.}, 239:493--521, 2003.

\bibitem[BB03b]{BauerBernard2003b}
M.~Bauer and D.~Bernard.
\newblock {SLE} martingales and the {Virasoro} algebra.
\newblock {\em Phys. Lett. B}, 557:309--316, 2003.

\bibitem[BB04]{BauerBernard2004a}
Michel Bauer and Denis Bernard.
\newblock Conformal transformations and the {SLE} partition function
  martingale.
\newblock {\em Ann. Henri Poincar{\'e}}, 5:289--326, 2004.

\bibitem[BBK05]{BauerBernardKytola2005}
M.~Bauer, D.~Bernard, and K.~Kyt{\"o}l{\"a}.
\newblock Multiple {Schramm-Loewner} evolutions and statistical mechanics
  martingales.
\newblock {\em J. Stat. Phys.}, 120:1125--1163, 2005.

\bibitem[BGLW05]{BettelheimGruzbergLudwigWiegmann2005}
E.~Bettelheim, I.~A. Gruzberg, A.~W.~W. Ludwig, and P.~Wiegmann.
\newblock Stochastic {Loewner} evolution for conformal field theories with
  {Lie} group symmetries.
\newblock {\em Phys. Rev. Lett.}, 95:251601, 2005.

\bibitem[BPZ84]{BelavinPolyakovZamolodchikov1984}
A.~A. Belavin, A.~M. Polyakov, and A.~B. Zamolodchikov.
\newblock Infinite conformal symmetry in two-dimensional quantum field theory.
\newblock {\em Nucl. Phys. B}, 241:333--380, 1984.

\bibitem[Car86]{Cardy1986}
J.~L. Cardy.
\newblock Effect of boundary conditions on the operator center of
  two-dimensional conformally invariant theories.
\newblock {\em Nucl. Phys. B}, 275:200--218, 1986.

\bibitem[Car89]{Cardy1989}
J.~L. Cardy.
\newblock Boundary conditions, fusion rules and the {Verlinde} formula.
\newblock {\em Nucl. Phys. B}, 324:581--596, 1989.

\bibitem[Car92]{Cardy1992}
J.~L. Cardy.
\newblock Critical percolation in finite geometries.
\newblock {\em J. Phys. A: Math. Gen.}, 25:L201--L206, 1992.

\bibitem[CDCH{\etalchar{+}}14]{ChelkakDuminil-CopinHonglerKemppainenSmirnov2014}
D.~Chelkak, H.~Duminil-Copin, C.~Hongler, A.~Kemppainen, and S.~Smirnov.
\newblock Convergence of {Ising} interfaces to {Schramm's SLE} curves.
\newblock {\em Comptes Rendus Mathematique}, 352:157--161, 2014.

\bibitem[DFMS97]{DiFrancescoMathieuSenechal1997}
Philippe Di~Francesco, Pierre Mathieu, and David S{\'e}n{\'e}chal.
\newblock {\em Conformal Field Theory}.
\newblock Graduate Texts in Contemporary Physics. Springer-Verlag New York,
  Inc., 1997.

\bibitem[Dub15a]{Dubedat2015b}
J.~Dub{\'e}dat.
\newblock {SLE} and {Virasoro} representations: Fusion.
\newblock {\em Commun. Math. Phys.}, 336:761--809, 2015.

\bibitem[Dub15b]{Dubedat2015a}
J.~Dub{\'e}dat.
\newblock {SLE} and {Virasoro} representations: Localization.
\newblock {\em Commun. Math. Phys.}, 336:695--760, 2015.

\bibitem[FBZ04]{FrenkelBen-Zvi2004}
E.~Frenkel and D.~Ben-Zvi.
\newblock {\em Vertex Algebras and Algebraic Curves}, volume~88 of {\em
  Mathematical Surveys and Monographs}.
\newblock American Methematical Society, 2nd edition, 2004.

\bibitem[FK80]{FrenkelKac1980}
I.~B. Frenkel and V.~G. Kac.
\newblock Basic representations of affine {Lie} algebras and dual resonance
  models.
\newblock {\em Invent. Math.}, 62:23--66, 1980.

\bibitem[FK04]{FriedrichKalkkinen2004}
R.~Friedrich and J.~Kalkkinen.
\newblock On conformal field theory and stochastic {Loewner} evolution.
\newblock {\em Nucl. Phys. B}, 687:279--302, 2004.

\bibitem[Fri04]{Friedrich2004}
R.~Friedrich.
\newblock On connections of conformal field theory and stochastic {Loewner}
  evolution, 2004.
\newblock arXiv:math-ph/0410029.

\bibitem[Fuk17]{Fukusumi2017}
Yoshiki Fukusumi.
\newblock Multiple {Schramm-Loewner} evolutions for coset {Wess-Zumino-Witten}
  models, 2017.
\newblock arXiv:1704.06006.

\bibitem[FZ92]{FrenkelZhu1992}
I.~B. Frenkel and Y.~Zhu.
\newblock Vertex operator algebras associated to representations of affine and
  {Virasoro} algebras.
\newblock {\em Duke Math. J.}, 66:123--168, 1992.

\bibitem[Kac98]{Kac1998}
V.~Kac.
\newblock {\em Vertex Algebras for Beginners}, volume~10 of {\em University
  Lecture Series}.
\newblock American Mathematical Society, Providence, RI, 2nd edition, 1998.

\bibitem[Kon03]{Kontsevich2003}
M.~Kontsevich.
\newblock {CFT, SLE} and phase boundaries, 2003.
\newblock Oberwolfach Arbeitstagung.

\bibitem[Kos18]{SK2018a}
S.~Koshida.
\newblock Local martingales associated with {Schramm-Loewner evolutions} with
  internal symmetry.
\newblock {\em J. Math. Phys.}, 59:101703, 2018.
\newblock arXiv:1803.06808.

\bibitem[Kyt07]{Kytola2007}
K.~Kyt{\"o}l{\"a}.
\newblock Virasoro module structure of local martingales of {SLE} variants.
\newblock {\em Rev. Math. Phys.}, 5:455--509, 2007.

\bibitem[Law04]{Lawler2004}
G.~F. Lawler.
\newblock An introduction to the stochastic {Loewner} evolution.
\newblock In {\em Random Walks and Geometry}. De Gruyter, 2004.

\bibitem[Law05]{Lawler2005}
Gregory~F. Lawler.
\newblock {\em Conformally Invariant Processes in the Plane}, volume 114 of
  {\em Mathematical Surveys and Monographs}.
\newblock American Mathematical Society, Providence, RI, 2005.

\bibitem[LR04]{LesageRasmussen2004}
F.~Lesage and J.~Rasmussen.
\newblock {SLE}-type growth processes and the {Yang-Lee} singularity.
\newblock {\em J. Math. Phys.}, 45:3040--3048, 2004.

\bibitem[MARR04]{Moghimi-AraghiRajabpourRouhani2004}
A.~Moghimi-Araghi, M.~A. Rajabpour, and S.~Rouhani.
\newblock Logarithmic conformal null vectors and {SLE}.
\newblock {\em Phys. Lett. B}, 600:298--301, 2004.

\bibitem[Naz12]{Nazarov2012}
A.~Nazarov.
\newblock {Schramm-Loewner} evolution martingales in coset conformal field
  theory.
\newblock {\em JETP Letters}, 96:90--93, 2012.

\bibitem[NR05]{NagiRasmussen2005}
J.~Nagi and J.~Rasmussen.
\newblock On stochastic evolutions and superconformal field theory.
\newblock {\em Nucl. Phys. B}, 704:475--489, 2005.

\bibitem[Ras04a]{Rasmussen2004b}
J.~Rasmussen.
\newblock Note on stochastic {L}{\"o}wner evolutions and logarithmic conformal
  field theory.
\newblock {\em J. Stat. Mech.}, page P09007, 2004.

\bibitem[Ras04b]{Rasmussen2004a}
J.~Rasmussen.
\newblock Stochastic evolutions in superspace and superconformal field theory.
\newblock {\em Lett. Math. Phys.}, 68:41--52, 2004.

\bibitem[Ras07]{Rasmussen2007}
J.~Rasmussen.
\newblock On {$SU(2)$ Wess-Zumino-Witten} models and stochastic evolutions.
\newblock {\em Afr. J. Math. Phys.}, 4:1--9, 2007.

\bibitem[RS05]{RohdeSchramm2005}
S.~Rohde and O.~Schramm.
\newblock Basic properties of {SLE}.
\newblock {\em Ann. Math.}, 161:883--924, 2005.

\bibitem[Sak13]{Sakai2013}
K.~Sakai.
\newblock Multiple {Schramm-Loewner} evolutions for conformal field theories
  with {Lie} algebra symmetries.
\newblock {\em Nucl. Phys. B}, 867:429--447, 2013.

\bibitem[Sch00]{Schramm2000}
O.~Schramm.
\newblock Scaling limits of loop-erased random walks and uniform spanning
  trees.
\newblock {\em Israel J. Math.}, 118:221--288, 2000.

\bibitem[Smi01]{Smirnov2001}
S.~Smirnov.
\newblock Critical percolation in the plane: conformal invariance, {Cardy's}
  formula, scaling limits.
\newblock {\em C. R. Acad. Sci. Paris}, 333:239--244, 2001.

\bibitem[Smi06]{Smirnov2006}
Stanislav Smirnov.
\newblock Towards conformal invariance of 2{D} lattice models.
\newblock In {\em Proceedings of the International Congress of Mathematicians
  (Madrid, August 22-30, 2006)}, pages 1421--1451. Eur. Math. Soc., Z{\"u}lich,
  2006.

\end{thebibliography}

\end{document}